\documentclass[reprint,twocolumn,amsmath,amssymb,aps,pra]{revtex4}

\usepackage{txfonts}
\usepackage{graphicx}
\usepackage{hyperref}
\usepackage{dcolumn}
\usepackage{bm}
\usepackage{textcomp}
\usepackage{setspace}
\usepackage{float}
\usepackage{indentfirst}
\usepackage{natbib}
\usepackage{subfigure}
\usepackage{booktabs}
\usepackage{multirow}
\usepackage{txfonts}
\usepackage{amsmath}
\usepackage{braket}
\usepackage{mathrsfs}
\usepackage{ulem}
\usepackage[titletoc]{appendix}

\allowdisplaybreaks
\makeatletter

\newcommand{\Rmnum}[1]{\expandafter\@slowromancap\romannumeral #1@}
\makeatother
\hypersetup{colorlinks=true,allcolors=blue,urlcolor=cyan,pdfstartview=Fit,breaklinks=true}

\begin{document}
	\title{ Controlling the $\mathcal{PT}$ Symmetry Breaking Threshold in Bipartite Lattice Systems with Floquet Topological Edge States}
	\author{Xinguang Li$^{1}$}
	\author{Hongzheng Wu$^{1}$}
	\author{Yangchun Zhao$^{1}$}
	\author{Jinpeng Xiao$^{2}$}
	\author{Yu Guo$^{3}$}
	\author{Lei Li$^{2}$}
	\author{Yajiang Chen$^{1}$}
	\author{Xiaobing Luo$^{1}$}
	\altaffiliation{Corresponding author: xiaobingluo2013@aliyun.com}
	\affiliation{$^{1}$Zhejiang Key Laboratory of Quantum State Control and Optical Field Manipulation, Department of Physics, Zhejiang Sci-Tech
		University, Hangzhou, 310018,  China}
	\affiliation{$^{2}$School of Mathematics and Physics, Jinggangshan University, Jinggangshan 343009, China}
	\affiliation{$^{3}$Hunan Provincial Key Laboratory of Flexible Electronic Materials Genome Engineering, School of Physics and Electronic Science,
		Changsha University of Science and Technology, Changsha 410114, China}

	\date{\today}
	\begin{abstract}
		We investigate the control of the parity-time ($\mathcal{PT}$)-symmetry breaking threshold in a periodically driven one-dimensional dimerized lattice with spatially symmetric gain and loss defects. We elucidate the contrasting roles played by Floquet topological edge states in determining the $\mathcal{PT}$ symmetry breaking threshold within the high- and low-frequency driving regimes. In the high-frequency regime, the participation of topological edge states in $\mathcal{PT}$ symmetry breaking is contingent upon the position of the $\mathcal{PT}$-symmetric defect pairs, whereas in the low-frequency regime, their participation is unconditional and independent of the defect pairs placement, resulting in a universal zero threshold.
		We establish a direct link between the symmetry-breaking threshold and how the spatial profile of the Floquet topological edge states evolves over one driving period. 
		We further demonstrate that lattices with an odd number of sites exhibit unique threshold patterns, in contrast to even-sized systems. Moreover, applying co-frequency periodic driving to the defect pairs, which preserves time-reversal symmetry, can significantly enhance the $\mathcal{PT}$ symmetry-breaking threshold.
		
	\end{abstract}
	\maketitle
	\section{Introduction}
	In recent years, research in topological physics has substantially advanced the frontiers of various physical systems.
	The characterization of topological phenomena fundamentally relies on topological invariants, which have been extensively investigated across diverse settings including non-Hermitian systems~\cite{kawabata2019symmetry,guo2022reentrant,rui2023making,tang2022experimental,long2022non,solnyshkov2021quantum,lin2023topological,PhysRevX.8.031079,song2019non,yao2018edge} and periodically driven systems~\cite{zhu2024dynamic,roy2017periodic}. For one-dimensional periodic system, the most commonly employed topological invariants are the Zak phase~\cite{zhu2018topological} and the winding number~\cite{asboth2014chiral}. These invariants can be mathematically derived to characterize topological systems and to predict the number of boundary states through the bulk-edge correspondence. Additionally, dynamical detection approach has been developed to directly measure these invariants, thereby providing direct insight into the topological nature of quantum systems~\cite{zhu2024dynamic,zhang2024dynamical}.

	As research on topological systems advances, the integration of non-Hermiticity into these frameworks has emerged as a natural progression. A central focus has been on the paradigmatic topological models, specifically the Aubry-André-Harper (AAH) model, which encompasses both diagonal AAH models~\cite{2014Aspects} and off-diagonal AAH models~\cite{wang2017quantum}, as well as the Su-Schrieffer-Heeger (SSH) model.
	The non-Hermitian extensions of the AAH and SSH models, incorporating gain and loss potentials, have been extensively studied~\cite{kawabata2019topological, liang2014pt, yuce2019topological, stegmaier2021topological, lang2018effects, turker2018pt, zeng2017anderson, pan2020interaction, wang2021quantum, bender1998real, su1979solitons, jin2017schrieffer, xu2020fate, garmon2021reservoir, yuce2018pt, wu2021topology, hu2023anti}, revealing a rich variety of novel phenomena.
	Furthermore, the introduction of non-reciprocal hopping into the SSH model has generated another wealth of theoretical results~\cite{geng2023separable,liu2025zak,li2021nonreciprocal}. These theoretical predictions have been substantiated by experiments, such as the verification of 
	$\mathcal{PT}$-symmetric topological interface states~\cite{weimann2017topologically}.
	The exploration of \(\mathcal{PT}\) symmetry has continued to deepen, with studies investigating its symmetry-breaking thresholds and topological states in greater detail~\cite{jin2017schrieffer,jin2018parity,blose2020floquet,harter2016mathcal}.
	Experimentally, $\mathcal{PT}$-symmetric gain and loss have been implemented across diverse platforms, including cold atom traps~\cite{creffield2007quantum,takasu2020pt}, photonic waveguides~\cite{rechtsman2013photonic}, optical lattices~\cite{longhi2014bound,kogel2019realization}, other optical devices and oscillators~\cite{sakhdari2018low,zhu2024all,dai2022dual,liu2021parity,liu2022narrow}, and electrical circuits~\cite{quijandria2018pt,purkayastha2020emergent,zeng2020topological}. Such experimental advancements have made it feasible to conduct in-depth theoretical studies on topological phases in non-Hermitian systems.
	Another key development in non-Hermitian physics is the non-Hermitian skin effect, where the bulk eigenstates become localized at the system's boundaries. This phenomenon has been widely discussed theoretically~\cite{li2022gain,kawabata2023entanglement,liang2022dynamic} and has also been realized in experiments~\cite{zhang2023electrical}.
	
	Floquet engineering serves as a powerful tool for dynamically controlling topological phases~\cite{zhou2023nonhermition,wu2020floquet,eckardt2015high,zhou2018recipe,kitagawa2010topological}.
	By tuning parameters such as the frequency or amplitude of a driving field, one can actively manipulate a system’s topological properties~\cite{PhysRevLett.109.106402,RevModPhys.91.015006,jotzu2014experimental,lindner2011floquet}. For instance, topologically protected boundary states can be selectively induced or eliminated at specific driving frequencies~\cite{nathan2015topological,peng2019floquet}. Furthermore, high-frequency driving has been used to engineer effective SSH models in diverse physical platforms, such as by periodically bending waveguides or shaking optical lattices in cold-atom systems~\cite{creffield2007quantum,zhu2018topological}.
	However, the interplay between $\mathcal{PT}$-symmetry breaking thresholds and topological boundary states in driven systems is not yet fully understood. This knowledge gap is especially pronounced in the low-frequency regime, which has received considerably less attention than the well-studied high-frequency limit.

	In this work, we demonstrate control over the parity-time  symmetry-breaking threshold in a non-Hermitian, dimerized lattice by engineering its Floquet topological phases with periodic driving, leveraging the properties of the topological edge states.
	Our results reveal a stark contrast in how topological boundary states affect the threshold in the high-frequency versus the low-frequency regime.
	In the high-frequency regime, whether the topological edge states participate in the $\mathcal{PT}$ symmetry breaking depends on the position of the $\mathcal{PT}$-symmetric defect pairs in the lattice. In contrast, in the low-frequency regime, the topological edge states always participate in the $\mathcal{PT}$ symmetry-breaking, regardless of the placement  of the $\mathcal{PT}$ defect pair, leading to a universal zero threshold.
	This distinction arises from the differing nature of the dynamical evolution of the spatial profile of the zero-energy topological edge states over one driving period in the low- and high-frequency regimes.
	Additionally, we find that the dependence of the threshold on the defect placement is sensitive to the parity of the number of lattice sites (i.e., whether it is odd or even).
	We also find that applying a double driving to both the lattice and the $\mathcal{PT}$-symmetric defect pairs causes the high-frequency zeroth-order effective Hamiltonian to behave like a Hermitian one, which, in turn, significantly enhances the $\mathcal{PT}$-symmetry breaking threshold.

	The paper is organized as follows. In Sec.~$\textcolor{blue}{\mathrm{II}}$, we present the model equation and discuss potential experimental implementations.
	In Sec.~\(\textcolor{blue}{\mathrm{III}}\), we demonstrate that the $\mathcal{PT}$ symmetry breaking threshold can be controlled by high-frequency driving, by inducing a Floquet topological phase, and connect this threshold to the spatial distribution of Floquet topological edge states. There are two mechanisms that induce \(\mathcal{PT}\)  symmetry breaking: Floquet topological edge states and bound states arising from \(\mathcal{PT}\)-symmetric defects. The involvement of these topological edge states in the \(\mathcal{PT}\) symmetry breaking process at high frequencies depends critically on the placement of the defect pairs.
	In Sec.~$\textcolor{blue}{\mathrm{IV}}$, we demonstrate that topological edge states are always involved in $\mathcal{PT}$ symmetry breaking at low frequencies. Their spatial profile oscillates periodically, aligning with the undriven case (which has zero components on even-numbered lattice sites and their reflection-symmetric partners) only when chiral symmetry is momentarily restored. This dynamic behavior is the key mechanism that causes the threshold to vanish.
	In Sec.~$\textcolor{blue}{\mathrm{V}}$, we show that zero-energy edge states persist across all driving parameters in odd-sized lattice, and further demonstrate that lattices with an odd number of sites exhibit distinct threshold patterns, in contrast to those with an even number of sites.
	In Sec.~$\textcolor{blue}{\mathrm{VI}}$, we find that applying co-frequency driving to defect pairs can render the zeroth-order approximation of the effective Hamiltonian Hermitian, leading to a significant enhancement of the $\mathcal{PT}$-symmetry breaking threshold.
	In Sec.~$\textcolor{blue}{\mathrm{VII}}$, we conclude the paper by summarizing the key findings of our study.

	\section{Model Equation}
\begin{figure*}[htbp]
		\centering
		\includegraphics[trim=0 0 0 0,clip,width=18cm]{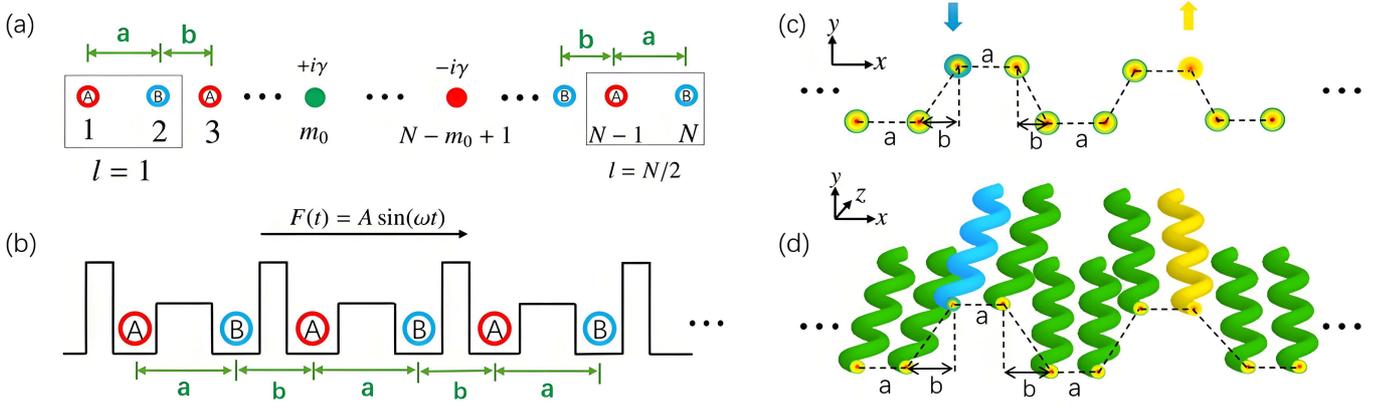}
		\caption{(a) and (b): Schematic of the realization of model~\eqref{con:1} in a cold-atom system. In (a), circles denote the sites of sublattices $A$ and $B$, each of which hosts a single state. The box highlights  a single unit cell of the system. The total number of chains, $N$, is even. The balanced gain and loss are located at site $m_0$ and its reflection-symmetric counterpart, site $N+1-m_0$. (b) Schematic of a bipartite lattice with alternating spacings $a$ and $b$. (c) and (d): Schematic for implementing model~\eqref{con:1} in a waveguide array. (c) Cross-sectional view in the $xy$-plane, depicting the bipartite lattice structure of the waveguide array. The horizontal center-to-center spacing alternates between $a$ and $b$.
			(d) Three-dimensional schematic showing the waveguides curving sinusoidally within the $xz$-plane as the light propagates along the $z$-axis. The vertical separation between waveguides in the $y$-direction alternates between a value of zero (aligned) and a fixed spacing.
		}
		\label{fig1}
	\end{figure*}

	We consider a single ultracold atom hopping on a one-dimensional bipartite lattice with an alternating $ABABAB$ structure, as shown in Figs. \ref{fig1}(a)-(b). In this lattice, the separation between adjacent $A$ and $B$ sites is asymmetric: the $A$-$B$ distance differs from the $B$-$A$ separation. One possible realization of this configuration is a chain of coupled double-well potentials. In this work, unless otherwise stated, we consider an $N$-site system with even $N$, incorporating a periodically oscillating potential that rises linearly along the lattice. Within this system, non-Hermitian gain and loss defects are positioned at reflection-symmetric sites $m_0$ and $N+1-m_0$. The Hamiltonian of the non-Hermitian extended chain is given by\cite{creffield2007quantum,jin2017schrieffer}
	\begin{equation}\label{con:1}
		H(t)=\tilde{H}(t)+H^{\gamma},
	\end{equation}
	with \begin{align*}
		\tilde{H}(t)&=\sum_{n=1}^{N-1} J(\left| n \right\rangle\langle n+1 |+\left| n+1 \right\rangle\langle n |)+A\sin(\omega t)\sum_{n=1}^{N} x_{n}\left| n \right\rangle\langle n |,\\
		H^{\gamma}&=i\gamma\left| m_0 \right\rangle\langle  m_0 | - i\gamma\left| N-m_0+1 \right\rangle\langle  N-m_0+1 |,
	\end{align*}
	where $\tilde{H}(t)$ describes the time-dependent Hermitian Hamiltonian of the system, with $A$ and $\omega$ parameterizing the amplitude and frequency of the shaken potential, and $H^{\gamma}$ describes reflection-symmetric gain-loss potentials. In Eq.~\eqref{con:1}, $\left| n \right\rangle$ represents the Wannier state localized at the $n$-th lattice site, and $
	J$ denotes the tunneling amplitude between nearest-neighbor sites. For simplicity, we assume equal coupling constants between adjacent sites, without loss of generality.
	For the bipartite lattice of Fig.~\ref{fig1}, the $x$-component of the location of lattice site $n$ is denoted $x_n$, which satisfies 
	\begin{equation}\label{con:2}
		x_{n+1}-x_n =
		\begin{cases}
			a & \ \ \ \ \ \ \ \ \ n=1 \mod 2 , \\
			b & \ \ \ \ \ \ \ \ \ n=0 \mod 2 .
		\end{cases}
	\end{equation} 
	
	An alternative realization of our model involves a photonic waveguide array, curved along the propagation direction of light ($z$ axis), as illustrated in Figs.~\ref{fig1}(c)-(d).	Along the propagation direction (\( z \)), the \( x \)-coordinates of waveguide centers vary periodically, while their \( y \)-coordinates remain constant. As shown in Figs.~\ref{fig1}(c)-(d), the structure consists of two layers of waveguide arrays along the $y$ axis. The center-to-center spacing between adjacent waveguides in the $x$ direction alternates between $a$ and $b$.
	The waveguide array includes one waveguide with gain (blue) and one  with loss (yellow), with all other waveguides being normal. The same evolution equation, governed by the Hamiltonian in Eq.~\eqref{con:1}, describes light propagation in the photonic waveguides shown in Figs.~\ref{fig1}(c)-(d) (see Appendix for details).

	\section{Control of the $\mathcal{PT}$-SYMMETRY BREAKING Threshold via High-Frequency Driving}
	In this section, we first focus on bipartite lattices to investigate how bound states or topological boundary states can be controlled by adjusting driving parameters in the high-frequency regime ($\omega \gg J$). To do this, we introduce the following time-dependent unitary transformation (rotation) operator
	\begin{equation}\label{con:3}
		\begin{aligned}
			\hat{S}(t)=&\text{exp}\left[-i\int A\sin(\omega t)\sum_{n}^{N} x_{n} \left| n \right\rangle\langle n |dt\right]\\
			=&\text{exp}\left[-i \frac{A}{\omega}\cos(\omega t)\sum_{n}^{N} x_{n} \left| n \right\rangle\langle n |\right].
		\end{aligned}
	\end{equation}
	Assuming \(x_{n+1} - x_n = a\) for odd \(n\) and \(x_{n+1} - x_n = b\) for even \(n\), and considering an even \(N\), this unitary transformation yields the new Hamiltonian as follows
	\begin{equation}\label{con:4}
		\begin{split}
			H^{\prime}(t)&=\hat{S}^{\dagger } (t) H(t) \hat{S} (t)-i\hat{S}^{\dagger } (t)\frac{\mathrm{d} \hat{S}(t) }{\mathrm{d} t}\\
			&=\tilde{H}^{\prime}(t)+H^{\gamma},
		\end{split}
	\end{equation}
	with the defect-free part,
	\begin{equation*}
		\begin{split}
			\tilde{H}^{\prime}(t)&=\hat{S}^{\dagger } (t) H(t) \hat{S} (t)-i\hat{S}^{\dagger } (t)\frac{\mathrm{d} \hat{S}(t) }{\mathrm{d} t}\\
			&=\sum_{l=1}^{N/2}\left(J\text{exp}\left[i\frac{A}{\omega}\cos(\omega t)a\right]\left|l,A\right\rangle\left\langle l,B\right|+h.c.\right)\\&+\sum_{l=1}^{N/2-1}\left(J\text{exp}\left[i\frac{A}{\omega}\cos(\omega t)b\right]\left|l+1,A\right\rangle\left\langle l,B\right|+h.c.\right).
		\end{split}
	\end{equation*}
	In Eq. \eqref{con:4}, for convenience we introduce \(\left|l,A\right\rangle\) and \(\left|l,B\right\rangle\) (with \(l \in \{1, 2, \dots, N/2\}\)), representing states where the particle occupies sublattice site \(A\) or \(B\), respectively, within unit cell \(l\), as illustrated in Fig. \ref{fig1}(a). The notation $``h.c."$ denotes the Hermitian conjugate.
	Before the transformation, the static part of Hamiltonian~\eqref{con:1} possesses  $\mathcal{PT}$  symmetry; however, the addition of the linear driving term breaks this symmetry. This transformation restores $\mathcal{PT}$ symmetry: $H^{\prime}(t)$ is invariant under the $\mathcal{PT}$ operation, which includes the transformations $\left|n\right\rangle \rightarrow \left|N-n+1\right\rangle$, $i \rightarrow -i$ and $t + t_0 \rightarrow -t + t_0$.
	
	Using the Floquet-Magnus expansion in the rotating frame, the effective Hamiltonian can be expressed as a series in powers of $1/\omega$,
	\begin{equation}\label{con:5}
		\begin{split}
			{H}_\text{{eff}}=\sum_{n=0}^{\infty} \frac{1}{\omega ^{n}}{H}_{\mathrm{eff}}^{(n)}. 
		\end{split}
	\end{equation}
	In the high-frequency regime, the system is accurately described by the leading-order term of the effective Hamiltonian expansion [Eq.~\eqref{con:5}]. This term corresponds to the time-averaged Hamiltonian in the rotating frame:
	\begin{equation}\label{con:6}
		\begin{split}
			H_{\mathrm{eff}}&\approx H_{\mathrm{eff}}^{(0)}=\tilde{H}_{\mathrm{eff}}^{(0)}+H^{\gamma}=\frac{1}{T}\int_{0}^{T}H^{\prime}(t)dt,\\
			\tilde{H}_{\mathrm{eff}}^{(0)}&=\frac{1}{T}\int_{0}^{T}\tilde{H}^{\prime}(t)dt\\
			&=\sum_{l=1}^{N/2} J_{1}\left(\left|l,A\right\rangle\left\langle l,B\right|+h.c.\right)\\
			&+\sum_{l=1}^{N/2-1} J_{2}\left(\left|l+1,A\right\rangle\left\langle l,B\right|+h.c.\right).
		\end{split}
	\end{equation}
	The hopping amplitudes are given by
	\begin{equation}\label{con:7}
		J_{1}=J \mathcal{J}_{0}\left(\frac{Aa}{\omega}\right),\ \ \
		J_{2}=J  \mathcal{J}_{0}\left(\frac{Ab}{\omega}\right),
	\end{equation}
	with \(\mathcal{J}_{0}\) denoting the zeroth-order Bessel function of the first kind.
	This leading-order effective Hamiltonian \(H_{\mathrm{eff}}^{(0)}\) realizes an SSH model incorporating one pair of \(\mathcal{PT}\)-symmetric defects. In this model: \(J_1\) represents intra-cell hopping; \(J_2\) represents inter-cell hopping. Through Floquet engineering, both hopping amplitudes can be dynamically controlled by tuning the dimensionless driving parameter \(A/\omega\). This creates a tunable SSH chain with engineered \(\mathcal{PT}\)-symmetric defects.
	
	\begin{figure}[htbp]
		\centering
		\includegraphics[trim=0 0 0 0,clip,width=9.2cm]{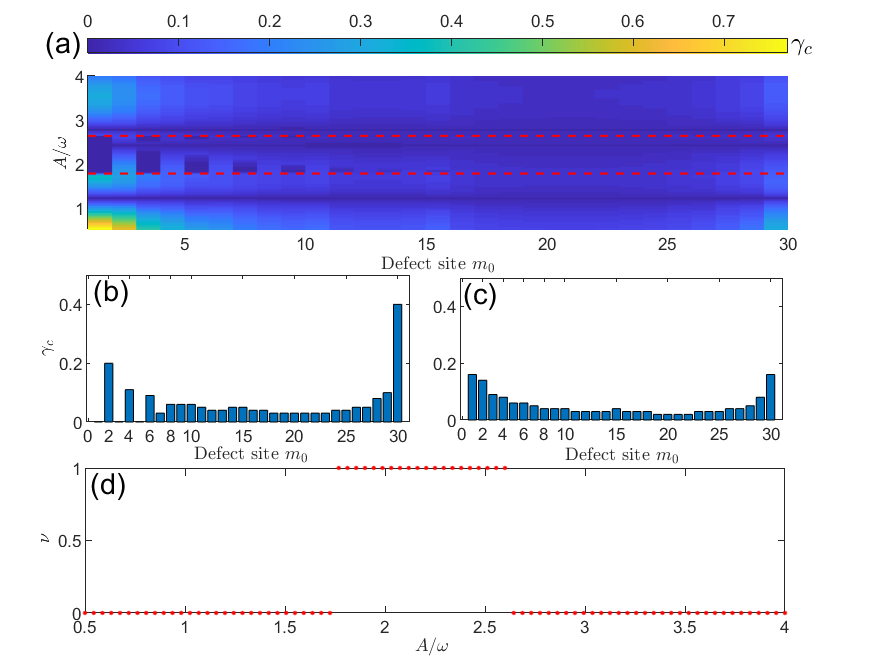}
		\caption{(a) The $\mathcal{PT}$-symmetry-breaking threshold, $\gamma_c$, plotted as a function of the gain-potential location, $m_0$, and the driving parameter, $A/\omega$. The area between the red dashed lines indicates the topological phase with winding number 1, as identified in Fig.~\ref{fig2}(d). (b) and (c): Threshold $\gamma_c$ versus \(m_0\) for slices of (a) at \(A/\omega=2\) (topologically non-trivial regime) and \(A/\omega=3\) (topologically trivial regime), respectively. (d) Winding number as a function of \(A/\omega\) for the same system without $\mathcal{PT}$-symmetric defects. Parameters used are \(\omega=20\), \(N=60\), \(J=1\), \(a=1\), and \(b=2\).
		}
		\label{fig2}
	\end{figure}
	
	As is known, the Hamiltonian after rotation, $H'(t)$ [see Eq.~\eqref{con:4}], is time-periodic and $\mathcal{PT}$ symmetric.
	Similar to the role of eigenenergies in static systems, periodically driven systems are characterized by quasienergies. 
	Generically, for a $\mathcal{PT}$-symmetric system, the quasienergy spectrum remains entirely real when $\gamma$ is below a critical threshold $\gamma_c$. However, when $\gamma$ exceeds $\gamma_c$, complex quasienergies emerge in complex-conjugate pairs. Therefore, $\gamma_c$ marks the phase transition between the $\mathcal{PT}$-broken and $\mathcal{PT}$-unbroken phases.
	In Fig.~\ref{fig2}(a), we show the $\mathcal{PT}$ symmetry breaking threshold strength $\gamma_c$ as a function of the gain location $m_0$ and the driving parameters $A/\omega$ for an $N = 60$ lattice.
	Numerically, the quasienergies can be evaluated by diagonalizing the Floquet operator \(U(\tau+T,\tau)\) in the real space, which satisfies the eigenvalue equation:     \(U(\tau+T,\tau)|\psi_{j}(\tau)\rangle=\mathrm{e}^{-\mathrm{i}\varepsilon_{j}T}|\psi_{j}(\tau)\rangle\), where $\varepsilon_{j}$ is called quasienergy; and $|\psi_{j}(\tau)\rangle=|u_{j}(\tau)\rangle \mathrm{e}^{-\mathrm{i}\varepsilon_{j}\tau}$, with $|u_{j}(\tau)\rangle =|u_{j}(\tau+T)\rangle$ and $T=2\pi/\omega$ representing the driving period.
	The Floquet operator is defined by the time-ordered exponential: \(U(\tau+T,\tau)=\mathbb{T}e^{-i\int_{\tau}^{\tau+T}H(t)dt}\), with \(\mathbb{T}\) denoting the time-ordering operator.
	When performing integrals involving exponential operators, one can substitute the original Hamiltonian $ H(t) $ with $ H'(t) $, as unitary transformations leave the energy spectrum unchanged.

	As shown in Fig.~\ref{fig2}(a), we place the gain site at $m_0$ and its conjugate loss site at the reflection-symmetric position $N-m_0+1$. For every configuration of $m_0$ and $A/\omega$, we determine the corresponding $\mathcal{PT}$ symmetry breaking threshold, the magnitude of which is indicated by the color map.
	We observe two distinct regions along the \(A/\omega\) axis, separated by two red dashed lines. In the region between the two lines, the threshold strength is zero for small odd values of \(m_0\) (where the gain and loss defects are close to the chain boundary). Outside these lines, this behavior is absent.
	For simplicity and without loss of generality, we set $a=1$ and $b=2$ in our model. As examples, we consider $A/\omega=2$ [Fig.~\ref{fig2}(b)] and $A/\omega=3$ [Fig.~\ref{fig2}(c)]. These choices highlight the distinct characteristics of the two regions. Specifically, when $A/\omega=2$, we clearly observe that the threshold $\gamma_c$ is zero for small odd gain locations $m_0$ (i.e., $m_0$ = 1, 3, 5), whereas for $A/\omega=3$, $\gamma_c$ is non-zero.
	
	In the absence of $\mathcal{PT}$-symmetric defects, the leading-order effective Hamiltonian $H_{\mathrm{eff}}^{(0)}$ exhibits spatial translational symmetry. Based on this symmetry, we transform $\tilde{H}_{\mathrm{eff}}^{(0)}$ into momentum space, resulting in the bulk momentum-space Hamiltonian as follows
	\begin{equation}\label{con:8}
		\begin{split}
			\tilde{H}_{\mathrm{eff}}^{(0)}(k)=\begin{pmatrix}
				0 & \Delta \\
				\Delta^{*} & 0
			\end{pmatrix},
		\end{split}
	\end{equation}
	with
	$\Delta=J_{1}\text{exp}(-ika)+J_{2}\text{exp}(-ikb)$.
	The corresponding bulk winding number is given by
	\begin{equation}\label{con:9}
		\nu=\frac{1}{2\pi i}\int_{BZ}dk\frac{d}{dk}\log \Delta,
	\end{equation}
	where $k$ is the Bloch wavenumber within the first Brillouin zone.
	The calculated winding number as a function of $A/\omega$ is shown in Fig.~\ref{fig2}(d). When the intracell hopping dominates the intercell hopping ($J_1 > J_2$), the winding number is $\nu = 0$. In the topological case, when $J_2 > J_1$, we have $\nu = 1$.
	We find that the two distinct regions observed in Fig.~\ref{fig2}(a), calculated from either the original Hamiltonian $H(t)$ or the rotated Hamiltonian $H^{\prime}(t)$ (unitary transformations preserving the spectrum), correspond one-to-one with the topological and non-topological regimes.
	This implies that the occurrence of a zero $\mathcal{PT}$ symmetry breaking threshold for small odd $m_0$ could be fundamentally related to topologically protected edge states.

\begin{figure}[htbp]
		\centering
		\includegraphics[trim=0 0 0 0,clip,width=9.2cm]{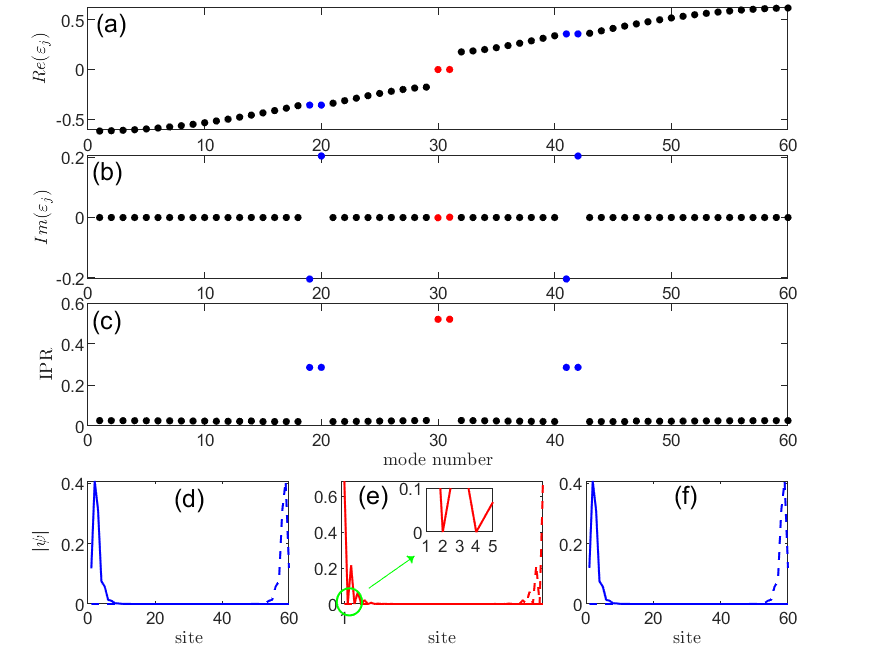}
		\caption{(a) Real parts of the quasi-energies, sorted in ascending order and plotted against their rank (mode number). (b) Corresponding imaginary parts of the quasi-energies, plotted against the mode number as in (a). (c) Inverse participation ratio (IPR) for each quasi-energy mode. (d) Spatial profiles of the defect-induced localized states corresponding to modes 19 (blue dashed) and 20 (blue solid). (e) Spatial profiles of the topologically protected zero-energy edge states corresponding to modes 30 (red dashed) and 31 (red solid). (f) Spatial profiles of a second pair of defect-induced localized states for modes 41 (blue dashed) and 42 (blue solid). The system parameters are $N=60$, $\omega = 20$, $A/\omega= 2$, $\gamma = 0.5$, $m_0 = 2$, $J=1$, $a = 1$, and $b = 2$.}
		
		\label{fig3}
	\end{figure}

	In the following, we investigate how topological edge states  influence $\mathcal{PT}$ symmetry breaking. We first select the topologically nontrivial regime \(A/\omega = 2\).  
	To investigate the role of topological edge states on the $\mathcal{PT}$ symmetry breaking threshold, this study will contrast the characteristics of edge states before and after symmetry breaking.
	As previously illustrated in Fig.~\ref{fig2}, within the topologically non-trivial region, for small \( m_0 \) (representing gain location), the threshold is zero if \( m_0 \) is odd, and non-zero if \( m_0 \) is even.
	In Fig.~\ref{fig3}, we present a specific example: reflection-symmetric gain-loss is placed at sites $m_0 = 2$ and $N-m_0+1 = 59$ (for $N = 60$) with $\gamma = 0.5$—a value not large yet exceeding the $\mathcal{PT}$ symmetry breaking threshold—to investigate the energy spectrum and distribution of bound states in the system under study.
	In Fig.~\ref{fig3}(a)-(b), we present the real and imaginary parts of the quasienergy spectrum, and in Fig.~\ref{fig3}(c) we show the inverse participation ratio (IPR). The IPR is defined as $\mathrm{IPR}=\langle u_{j} ^{2}(\tau)|u_{j}^{2}(\tau)\rangle/\langle u_{j}(\tau)|u_{j}(\tau)\rangle^{2}$  for the Floquet state $|u_{j}(\tau)\rangle=|u_{j}(\tau+T)\rangle$ corresponding to quasienergy $\varepsilon_{j}$, where the IPR is non-zero for localized modes and approximately 0 for extended states.
	The mode number on the horizontal axis is ordered according to increasing values of the real part of quasienergy.
	Our findings indicate that a pair of degenerate zero-energy topological edge states (marked in red) persist and remain unaffected, even with the introduction of non-Hermiticity and broken $\mathcal{PT}$ symmetry.
	Additionally, we identified two pairs of non-zero-energy bound states. These states possess positive (negative) imaginary eigenvalues and are centered at the gain (loss) site. One pair, corresponding to mode numbers 19 (dashed line, localized near the loss site) and 20 (solid line, localized near the gain site), is shown in Fig.~\ref{fig3}(d). The other pair, corresponding to mode numbers 41 (dashed line, localized near the loss site) and 42 (solid line, localized near the gain site), is also depicted in Fig.~\ref{fig3}(f).
	As shown in Fig.~\ref{fig3}(e), these two zero-energy topological edge states are localized on the left and right sides of the lattice, respectively.
	Upon closer inspection [see the inset in Fig.~\ref{fig3}(e)], it is found that the topologically protected edge state exponentially localized at the left edge has vanishing weight on the $B$ sublattice (even-$n$ sites), while its counterpart localized at the right edge has vanishing weight on the $A$ sublattice (odd-$n$ sites).
	That is, when \(m_0\) is even, these topological edge states exhibit no weight on the \(\mathcal{PT}\)-symmetric defect sites and thus  do not participate in \(\mathcal{PT}\)-symmetry breaking. Instead, the breaking originates from the bound states localized near the gain or loss sites, induced by the non-Hermitian \(\mathcal{PT}\)-symmetric defects.
	
	\begin{figure}[htbp]
		\centering
		\includegraphics[trim=0 0 0 0,clip,width=9.2cm]{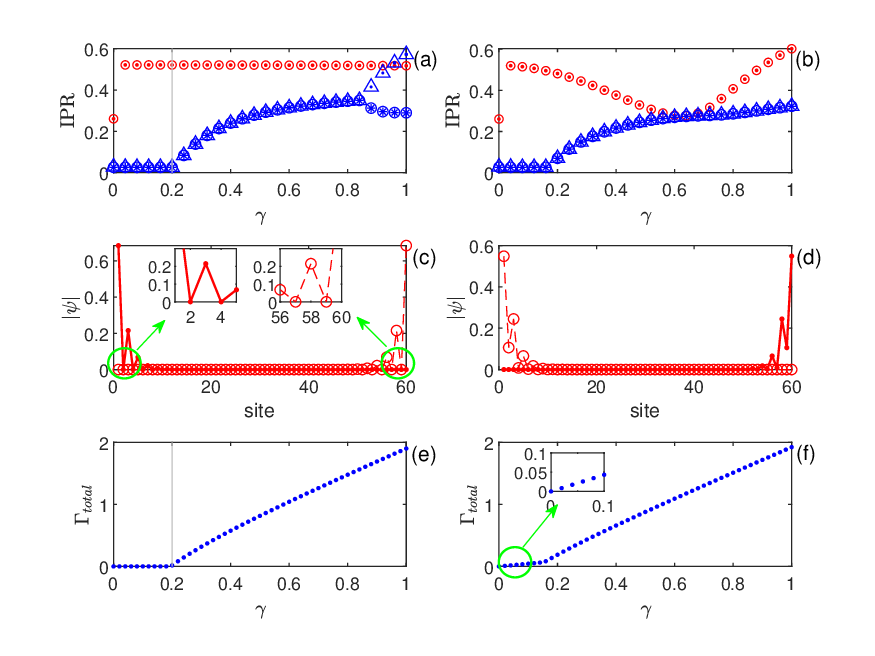}
		\caption{Top row: inverse participation ratio (IPR) as a function of the gain/loss parameter $\gamma$ for the topologically nontrivial regime ($A/\omega = 2$) with (a) $m_0 = 2$ and (b) $m_0 = 3$. In both panels, the two topological edge states are highlighted by red circles and dots, while the remaining four bound states induced by \(\mathcal{PT}\)-symmetric defects are shown in blue with various markers (triangles, dots, circles, and stars). Middle row: spatial profiles of a pair of degenerate topological edge states at \(\gamma = 0.4\), shown for the topologically nontrivial regime (\(A/\omega = 2\)) with (c) \(m_0 = 2\) and (d) \(m_0 = 3\). Bottom row: (e) and (f) show sum of the absolute values of the imaginary parts of all quasi-energies [see $\Gamma_{\text{total}}$ in Eq.~\eqref{con:10}], for the same parameters as (a) and (b), respectively. Other parameters: \(\omega = 20\), $N=60$, $J=1$, $a=1$ and $b=2$.}   
		\label{fig4}
	\end{figure}
	For a clearer elucidation of the relationship between topological edge states and \(\mathcal{PT}\)-symmetric defects-induced bound states with respect to the \(\mathcal{PT}\) symmetry breaking threshold, we investigate two specific cases within the topological region: $m_0 = 2$ and $m_0 = 3$. The corresponding results are presented in the left and right columns of Fig.~\ref{fig4}.
	In Figs.~\ref{fig4}(a)-(b), the red circles and dots represent the IPRs of the two topologically protected edge states, while the blue markers correspond to the IPRs of defect-induced localized bound states.
	When the defect strength $\gamma = 0$, the two zero-energy topological edge states are odd and even superpositions of states localized exponentially on the left and right edges. 
	When a defect pair is introduced, these two zero-energy topological edge states become localized at either the left or the right edge, causing a sudden change (or jump) in the IPR [see red markers in Figs.~\ref{fig4}(a)-(b)].
	Figures~\ref{fig4}(c) and (d) show the profiles of the boundary states for \( m_0 = 2 \) and \( m_0 = 3 \) respectively, at \( \gamma=0.4 \). 
	For the former (\( m_0 = 2 \)), neither the left- nor right-edge topological edge states have components at the defect sites, and thus are not involved in the \(\mathcal{PT}\) symmetry breaking. Therefore, their IPR is unaffected by the defect strength [see the red markers in Fig.~\ref{fig4}(a)].
	In contrast, for the latter (\( m_0 = 3 \)), the topological boundary state at \( \gamma = 0 \) already has components at the lattice sites where a defect pair is introduced. As a result, while introducing the defect preserves the boundary state, its profile and IPR undergo significant changes [see Figs.~\ref{fig4}(b) and (d)].
	However, regardless of defect placement location, the defect-induced bound states [see the blue markers in Figs.~\ref{fig4}(a)-(b)] only emerge above a finite threshold of defect strength.

	To precisely identify the $\mathcal{PT}$ symmetry breaking threshold, we numerically compute the sum of the absolute values of the
	imaginary parts of all quasienergies [see Figs.~\ref{fig4}(e)-(f)],
	\begin{equation}\label{con:10}
		\Gamma_{\text{total}} = \sum_{j} |\text{Im}(\varepsilon_j)|.
	\end{equation}
	The onset of non-zero \(\Gamma_{\text{total}}\) defines the \(\mathcal{PT}\) symmetry breaking threshold.
	When a gain-loss defect is placed on the $m_0$ site and its symmetric reflection site $N-m_0+1$, and when $m_0$ represents an even number (specifically $m_0=2$), the topological boundary state has no components at these defect sites, and it does not participate in the $\mathcal{PT}$ symmetry breaking. Consequently, the $\mathcal{PT}$ symmetry breaking arises from a bound state induced by the $\mathcal{PT}$-symmetric defects. In this case, the $\mathcal{PT}$ symmetry breaking threshold (marked by the gray line) corresponds one-to-one with the threshold of the defect-induced bound state [see Figs.~\ref{fig4}(a) and (e)].
	However, for small odd $m_0$, topologically induced boundary states have components at the defect lattice sites, which consequently leads to the $\mathcal{PT}$ symmetry breaking threshold being zero [see Fig.~\ref{fig4}(f)].
	We note that for larger odd $m_0$, the probabilities of the topologically induced edge states decay to zero at the defect pair sites. In such configurations, $\mathcal{PT}$ symmetry breaking is independent of the emergence of the edge states, and thus the $\mathcal{PT}$ symmetry breaking threshold is non-zero [as previously illustrated in Fig.~\ref{fig2}(b), see, for example, $ m_0=7 $], corresponding to the threshold for the appearance of defect-induced bound states.
	
	So far, it has become clear that within the high-frequency-induced topologically non-trivial regimes (as shown in Fig.~\ref{fig4}), the driving-induced topological edge states play contrasting roles in $\mathcal{PT}$-symmetry breaking depending on the defect location \( m_0 \). For even \( m_0 \), they do not contribute to the symmetry breaking, yielding a finite threshold. However, for small odd \( m_0 \), they actively drive the breaking, leading to a threshold of zero.
	Additionally, in the topologically trivial regions [refer to Fig.~\ref{fig2}(c)], there are no driving-induced topological edge states. Therefore, regardless of whether $m_0$ is odd or even, the $\mathcal{PT}$ symmetry breaking threshold is non-zero.
	
	\section{Control of the $\mathcal{PT}$-SYMMETRY BREAKING Threshold via Low-Frequency Driving}
	\begin{figure*}[!htbp]
		\centering
		\includegraphics[trim=0 0 0 0,clip,width=14cm]{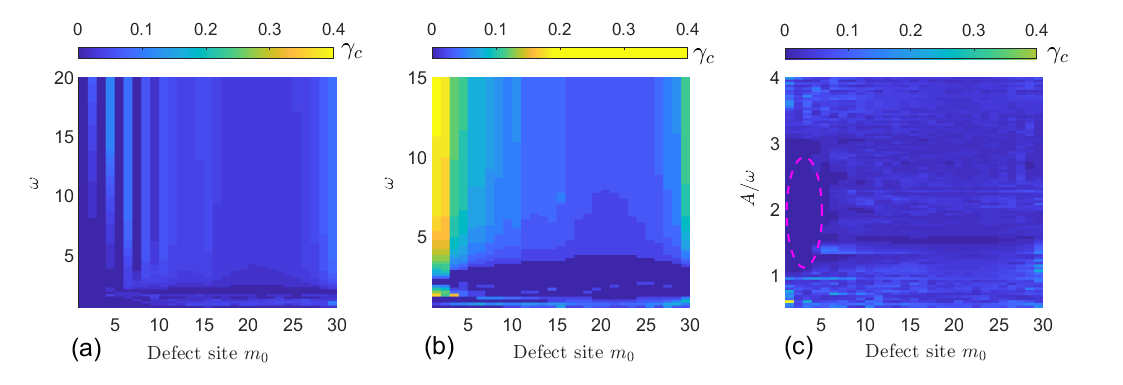}
		\caption{$\mathcal{PT}$ symmetry breaking threshold as a function of driving frequency $\omega$ and defect location $m_0$ for different driving amplitudes: (a) $A/\omega = 2$ and (b) $A/\omega = 1.4$. (c) Threshold as a function of the amplitude-to-frequency ratio $A/\omega$ and defect location $m_0$ for a fixed low frequency $\omega = 1$. The region enclosed by the pink dashed lines indicates where the threshold is strongly suppressed, approaching zero. Other parameters: $N = 60$, $J=1$, $a=1$, $b=2$.}
		\label{fig5}
	\end{figure*}
	
	In previous discussions, we focused on high-frequency regimes (specifically with \(\omega = 20\)), where the value of \(A/\omega\) determines the existence of topological edge states. These edge states significantly influence the system's \(\mathcal{PT}\) symmetry breaking threshold. This work further investigates whether similar topological edge states exist in low-frequency regimes and how they affect \(\mathcal{PT}\) symmetry breaking threshold.
	
	To address the aforementioned goals, Figs.~\ref{fig5}(a)-(b) show how the $\mathcal{PT}$ symmetry breaking thresholds vary with decreasing frequency for the two cases: $A/\omega = 2$ (topological phase at high frequencies) and $A/\omega = 1.4$ (non-topological phase at high frequencies).
	These calculations for the thresholds are performed for each configuration of defect location $m_0$, using the same steps described in Fig.~\ref{fig2}(a).
	As expected, at high frequencies, the threshold strength $\gamma_c$ 
	exhibits an alternating pattern of zero and non-zero values for small $m_0$ in Fig.~\ref{fig5}(a) (corresponding to $A/\omega=2$), whereas for $A/\omega=1.4$, $\gamma_c$ is non-zero for all $m_0$, as shown in Fig.~\ref{fig5}(b).
	However, irrespective of the specific configuration, in the low-frequency regime (for instance, $\omega = 1$), we observe that if the defect pairs are placed not sufficiently far from the lattice edges, the threshold is essentially zero.
	In Fig.~\ref{fig5}(c), we present the \(\mathcal{PT}\) symmetry breaking threshold strength $\gamma_c$ as a function of $A/\omega$ and $m_0$, with $\omega$ fixed at 1.
	Particularly noteworthy is the region marked by pink dashed circles in Fig.~\ref{fig5}(c), which exhibits a zero threshold. This significantly deviates from the high-frequency behavior observed in Fig.~\ref{fig2}(a), highlighting the fundamental difference between the low-frequency and  the high-frequency regime.
	As we will elaborate on later, this difference is due to the unique properties of the zero-energy topological edge states induced by low-frequency driving.
	
	Unlike in the high-frequency region where the Floquet-Magnus expansion permits approximating the time-periodic Hamiltonian by its zeroth-order time-independent Hamiltonian, this approximation fails in the low-frequency region.
	To define proper topological invariants, we need to consider the bulk Hamiltonian free of defects and with spatial translational symmetry.
	To this end, we write $\tilde{H}^{\prime}(t)$ [defined as $H^{\prime}(t)$ with $\gamma=0$; see Eq.~\eqref{con:4}] in momentum space, 
	\begin{equation}\label{con:11}
		\tilde{H}^{\prime}(k,t)=
		\begin{pmatrix}
			0 & \beth \\
			\beth^* & 0
		\end{pmatrix}=h_x(k,t)\sigma_x+h_y(k,t)\sigma_y,
	\end{equation}
	with
	\begin{align*}
		\beth&=J\text{exp}\left[i\frac{A}{\omega}\cos(\omega t)a\right]\text{exp}(ika)\\
		&+J\text{exp}\left[-i\frac{A}{\omega}\cos(\omega t)b\right]\text{exp}(-ikb),\\
		h_{x}(k,t)&=J\cos[\frac{A}{\omega}\cos(\omega t)a+ka]\\&+J\cos[\frac{A}{\omega}\cos(\omega t)b+kb],\\
		h_{y}(k,t)&=-J\sin[\frac{A}{\omega}\cos(\omega t)a+ka]\\&+J\sin[\frac{A}{\omega}\cos(\omega t)b+kb].
	\end{align*}
	Based on the bulk momentum-space Hamiltonian \eqref{con:11}, the time evolution operator over one period is defined by
	\begin{equation}\label{con:12}
		\tilde{U}(k,\tau)=\mathbb{T}\text{exp}\left[-i\int_{\tau}^{\tau+T}\tilde{H}^{\prime}(k,t)dt\right].
	\end{equation}
	Here, $\tilde{U}(k,\tau)$ is the time-evolution operator in momentum space mapping the state at time $\tau$ to the state at time $\tau+T$, where $\tau$ is the initial time.
	The corresponding Floquet Hamiltonian is defined by $\tilde{H}_{\mathrm{eff}}(k,\tau)\equiv i\log \tilde{U}(k,\tau)/T$, with parametric dependence on the initial time $\tau$.
	
	The Floquet modes $|\tilde{u}_j(k,\tau)\rangle$ and their quasienergies $\tilde{\varepsilon}_j(k)$ can be obtained from the diagonalization of $\tilde{H}_{\mathrm{eff}}(k,\tau)$,
	\begin{equation}\label{con:13}
		\tilde{H}_{\mathrm{eff}}(k,\tau)\left|\tilde{u}_j(k,\tau) \right\rangle=\tilde{\varepsilon}_j(k) \left|\tilde{u}_j(k,\tau) \right\rangle, \ \ (j=\pm).
	\end{equation}
	Thus, the Floquet Hamiltonian $\tilde{H}_{\mathrm{eff}}(k,\tau)$ can be numerically expressed in terms of Pauli matrices:
	\begin{equation}\label{con:14}
		\begin{aligned}
			\tilde{H
			}_{\mathrm{eff}}(k,\tau) & =h_x^{\mathrm{eff}}(k,\tau)\sigma_x+h_y^{\mathrm{eff}}(k,\tau)\sigma_y+h_z^{\mathrm{eff}}(k,\tau)\sigma_z, \\
			h_x^{\mathrm{eff}}(k,\tau) & =\tilde{\varepsilon}_+(k)\langle \tilde{u}_+(k,\tau)|\sigma_x|\tilde{u}_+(k,\tau)\rangle, \\
			h_y^{\mathrm{eff}}(k,\tau) & =\tilde{\varepsilon}_+(k)\langle \tilde{u}_+(k,\tau)|\sigma_y|\tilde{u}_+(k,\tau)\rangle, \\
			h_z^{\mathrm{eff}}(k,\tau) & =\tilde{\varepsilon}_+(k)\langle \tilde{u}_+(k,\tau)|\sigma_z|\tilde{u}_+(k,\tau)\rangle.
		\end{aligned}
	\end{equation}
	It is noted that the Floquet Hamiltonians corresponding to any two different initial times can be related by a unitary transformation. Therefore, the quasienergy spectrum is independent of the choice of the initial time $\tau$.
	
	Chiral symmetry plays a key role in determining the  topologically-nontrivial phase by utilizing the winding number within the two-band Hamiltonian~\cite{zhu2024dynamic}.
	The Floquet Hamiltonian does not necessarily inherit chiral symmetry from the instantaneous Hamiltonian $\tilde{H}^{\prime}(k, t)$, and it may lose its original chiral symmetry due to the existence of Floquet band couplings in the low-frequency regime.
	Thus, ensuring chiral symmetry in a periodically driven system requires identifying an initial time $\tau$ where the corresponding Floquet Hamiltonian exhibits chiral symmetry.
	\begin{figure}[htbp]
		\centering
		\includegraphics[trim=0 0 0 0,clip,width=9.2cm]{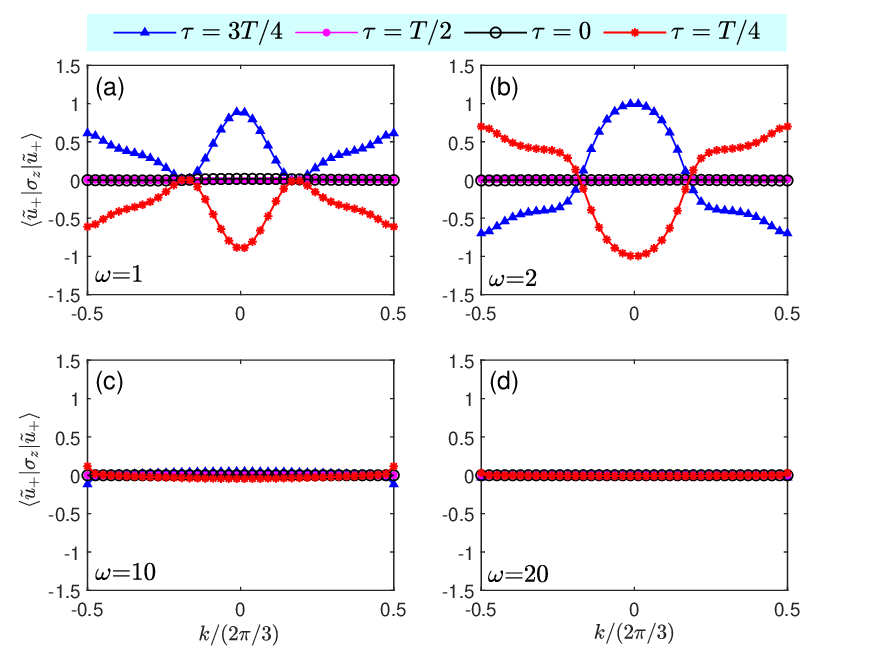}
		\caption{The expectation value $\langle \tilde{u}_+(k,\tau)|\sigma_z|\tilde{u}_+(k,\tau)\rangle$ [defined in Eq.~\eqref{con:14}] as a function of quasi-momentum $k$ for different driving frequencies, evaluated at different initial times $\tau$. From (a) to (d), the driving frequencies are 1, 2, 10, and 20, respectively. Other parameters: $J=1$, $A=2$, $a=1$ and $b=2$.}
		\label{fig6}
	\end{figure}
	
	In Fig.~\ref{fig6}, we plot $\langle \tilde{u}_+(k,\tau)|\sigma_z|\tilde{u}_+(k,\tau)\rangle$  as a function of quasimomentum $k$ across the first Brillouin zone for various initial time $\tau$ and driving frequencies $\omega$.
	When the initial time is chosen as $\tau = 0$ or $T/2$, the expectation value $\langle \tilde{u}_+(k,\tau) | \sigma_z | \tilde{u}_+(k,\tau) \rangle$ vanishes for all quasimomenta $k$, irrespective of changes in the driving frequency $\omega$, as illustrated by the black and cyan lines in Figs.~\ref{fig6}(a)-(d).
	However, in the low-frequency region, when we choose the initial time $\tau = 3T/4, T/4$, the expectation value $\langle \tilde{u}_+(k,\tau) | \sigma_z | \tilde{u}_+(k,\tau) \rangle$ becomes sensitive to changes in the quasi-momentum $k$, as shown in Figs.~\ref{fig6}(a)-(b).
	To restore the  chiral symmetry of the Floquet Hamiltonian $\tilde{H}_{\mathrm{eff}}(k,\tau)$, one effective approach is to increase the driving frequency \(\omega\). This increase reduces the coupling between Floquet bands, which in turn diminishes the sensitivity of the expectation value $\langle\tilde{u}_+(k,\tau)|\sigma_z|\tilde{u}_+(k,\tau)\rangle$ to changes in the initial time, as illustrated in Figs.~\ref{fig6}(c)–(d). Furthermore, when the driving frequency becomes sufficiently high, the Floquet Hamiltonian consistently retains its original chiral symmetry, irrespective of variations in the initial times, as demonstrated in Fig.~\ref{fig6}(d).
	
	At low frequencies, it is shown that the chiral symmetry is generally broken by the periodic driving, namely,
	$\sigma_z\tilde{H}_{\mathrm{eff}}(k,\tau)\sigma_z^{-1}\neq-\tilde{H
	}_{\mathrm{eff}}(k,\tau)$.
	If $\langle u_+(k,\tau)|\sigma_z|u_+(k,\tau)\rangle$ is zero for all quasi-momenta $k$, this means that the Floquet Hamiltonian $\tilde{H}_\mathrm{eff}(k,\tau)$ has chiral symmetry. Then, we can define proper topological invariants in our periodically driven system in a similar manner as the static system.
	In this case, the Floquet Hamiltonians in momentum space can be written in a block off-diagonal form,
	\begin{equation}\label{con:15}
		\tilde{H}_{\mathrm{eff}}(k,\tau)=\begin{pmatrix}
			0 & h^{\mathrm{eff}}(k,\tau) \\
			h^{{\mathrm{eff}*}}(k,\tau) & 0\end{pmatrix},
	\end{equation}
	with
	\begin{equation}\label{con:16}
		h^{\mathrm{eff}}(k,\tau)=h_x^{\mathrm{eff}}(k,\tau)-ih_y^{\mathrm{eff}}(k,\tau).
	\end{equation}
	
	In our periodically driven system, there exist two different initial times, \(\tau_{1}=0\) and \(\tau_{2}=T/2\), such that the Floquet Hamiltonians \(\tilde{H}(k,\tau_{1})\) and \(\tilde{H}(k,\tau_{2})\) both possess chiral symmetry.
	Thus, the two topological invariants (winding numbers) can be assigned as follows:
	\begin{equation}\label{con:17}
		\nu_{j}=\frac{1}{2\pi i}\int_{BZ}dk\frac{d}{dk}\ln [h^{\mathrm{eff}}(k,\tau_{j})]\ \ \ \ \ \ \ \ \ (j=1,2).
	\end{equation}
	
\begin{figure}[htbp]
		\centering
		\includegraphics[trim=0 0 0 0,clip,width=9.2cm]{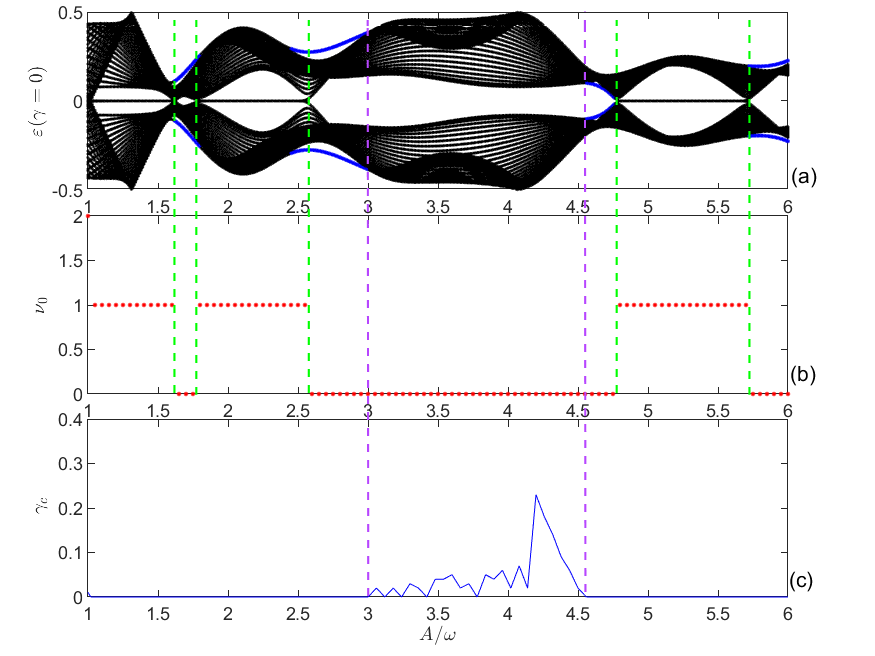}
		\caption{(a) Quasi-energy spectrum for an open chain (\(N=60\)) at a low driving frequency (\(\omega = 1\)), without \(\mathcal{PT}\)-symmetric defects, calculated via real-space Hamiltonian diagonalization. The non-zero energy gap boundary states are highlighted in blue. (b) The corresponding zero-mode winding number, computed using the same parameters as in (a). (c) The threshold value $\gamma_c$ plotted against \(A/\omega\) for a defect pairs positioned at \(m_0 = 2\) and its reflection-symmetric site. Other parameters: \(N=60\), \(J=1\), \(a=1\), and \(b=2\).}
		\label{fig7}
	\end{figure}

	According to Ref.~\cite{asboth2014chiral}, the winding numbers $\nu_1$ and $\nu_2$ can be combined to obtain the bulk topological invariants controlling the number of end states,
	\begin{equation}\label{con:18}
		\begin{aligned}
			\nu_0=\frac{\nu_{1}+\nu_{2}}{2},\quad\nu_\pi=\frac{\nu_{1}-\nu_{2}}{2},
		\end{aligned}
	\end{equation}
	where $\nu_{0}$ and $\nu_{\pi}$ represent the edge state winding numbers at quasienergies  0  (0-mode) and $ |\varepsilon| = \pi/T$ ($\pi$-mode), respectively,
	
	In Fig.~\ref{fig7}(a), we present the quasienergy spectrum of the original Hamiltonian \eqref{con:1} with \(\gamma=0\) at \(\omega=1\), under the open boundary condition with \(N=60\).
	As illustrated in Fig.~\ref{fig7}(b), the Floquet topological phases of the quasienergy spectrum under the open-boundary condition are well characterized by the winding numbers.
	In Fig.~\ref{fig7}(b) we plot the winding numbers of the 0-mode, which, as the bulk topological invariant under periodic boundary conditions, demonstrate a one-to-one correspondence with the spectrum under open boundary condition, thus establishing the exact bulk-edge correspondence.
	From our numerical computations, we observed $\nu_{1} = \nu_{2}$, and consequently, no $\pi$-mode edge states appear.
	In Fig.~\ref{fig7}(a), in addition to the zero-energy topological edge state, non-zero-energy gap states marked with blue lines are observed. Their presence is attributed to the creation of virtual defects at the boundary induced by the periodic driving \cite{zhu2018topological,garanovich2008defect}.
	
	To examine the link between the boundary states and the $\mathcal{PT}$ symmetry breaking threshold, we plot the threshold as a function of the driving parameter $A/\omega$ in Fig.~\ref{fig7}(c). The calculations are performed using the same parameters as in Fig.~\ref{fig7}(a), except for the introduction of defect pairs at positions $m_0=2$ and $N - m _0+ 1$.
	We observe that whenever boundary states appear, regardless of whether they are topological or non-topological, the $\mathcal{PT}$ symmetry breaking threshold is always 0. This indicates that the boundary states participate in the $\mathcal{PT}$ symmetry breaking. This behavior differs from the high-frequency scenario, where, when $m_0$ is even, topological edge states have no component at the defect pair locations and thus do not participate in the $\mathcal{PT}$ symmetry breaking.
	
	In the low-frequency regime,  the occurrence  of topological edge states invariably results in a $\mathcal{PT}$ symmetry breaking threshold of zero, regardless of defect location—a behavior quite different from the high-frequency case.
	To explain why zero-energy topological edge states play completely different roles in $\mathcal{PT}$ symmetry breaking in the high-frequency versus low-frequency regimes, we need to examine the spatial probability distribution of these eigenstates across the lattice sites.
	To address this, we analyze the sum of the spatial probability density for the zero-energy topological edge state at even sites in the left half of the chain and their reflection-symmetric counterparts in the right half, defining $\Sigma_P$ as:
	\begin{equation}\label{con:19}
		\Sigma_{P}=\sum_{m_0 = 2,4\ldots N/2}(P_{m_0}+P_{N-m_0+1}),
	\end{equation}
	where \( P_n \) denotes the probability density at site \( n \).
	
	We initialize the system in one of the two degenerate zero-energy eigenstates of the Floquet Hamiltonian and evolve it over one driving period, recording the sum of the modulus squared of the wave function at the specified sites at each time.
	At \(t=0\), the Floquet Hamiltonian exhibits chiral symmetry in both high- and low-frequency regimes. Under this symmetry, the zero-energy topological edge state displays vanishing probability density at all even-indexed sites in the left half of the chain and their reflection-symmetric counterparts in the right half. 
	From the Floquet Hamiltonian in real space (for the defect-free case), the time evolution of a quantum state initially prepared in a Floquet mode is given by: $|\psi(t)\rangle = \mathrm{e}^{-\mathrm{i}\tilde{\varepsilon}_{j}t}|\tilde{u}_{j}(t)\rangle$ where $|\tilde{u}_{j}(t)\rangle$ are the eigenstates of  $\tilde{H}_\mathrm{eff}(t)$  in real space, with corresponding quasienergies $\tilde{\varepsilon}_{j}$. 
	
	\begin{figure}[htbp]
		\centering
		\includegraphics[trim=0 0 0 0,clip,width=9.2cm]{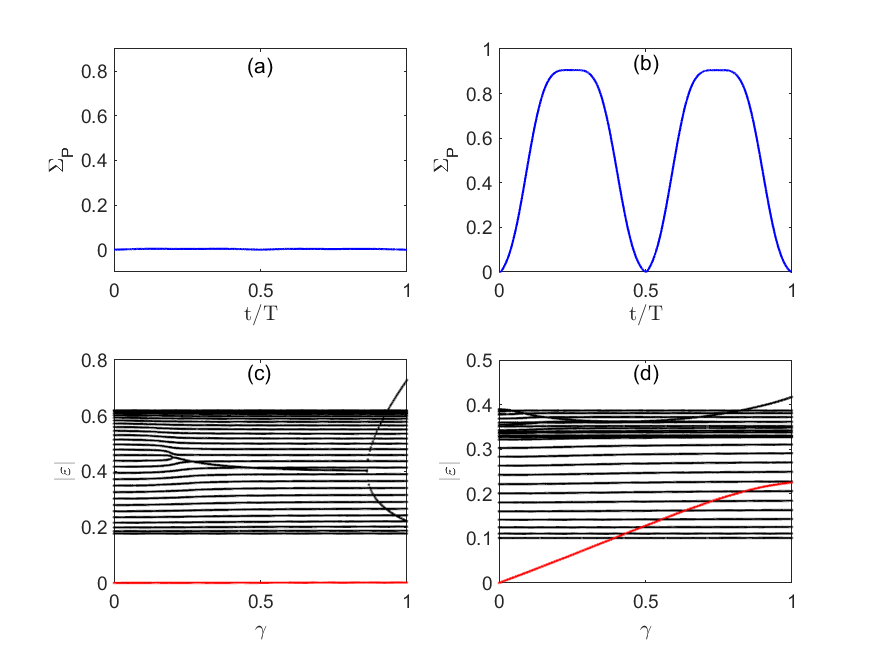}
		\caption{Top row: evolution of \(\Sigma_{P}\) [defined by Eq.~\eqref{con:19}] for the topological edge states over one driving period in a defect-free system: (a) high-frequency regime (\(\omega = 20\)) and (b) low-frequency regime (\(\omega = 1\)). Bottom row: magnitude of the quasi-energies versus \(\gamma\) for high frequency (c) \(\omega = 20\)  and low frequency (d) \(\omega = 1\) , with defects placed at \(m_0 = 2\) and its reflection-symmetric site. In both (c) and (d), the red lines represent how the magnitude of quasi-energies related to topological edge states changes with $\gamma$. Other parameters: $N=60$, $A/\omega=2$, $J=1$, $a=1$ and $b=2$.}
		\label{fig8}
	\end{figure}
	
	From Fig.~\ref{fig8}(a), we observe that in high-frequency regimes, \(\Sigma_{P}\) remains identically zero throughout the evolution. This indicates that the population distribution of the zero-energy Floquet mode experiences negligible change during the evolution. This explains why, in the high-frequency regime, the Floquet Hamiltonian can be described by the leading-order time-independent Hamiltonian.
	In contrast, in the low-frequency regime, \(\Sigma_{P}\) exhibits periodic oscillations, being zero only at integer and half-integer periods, as shown in Fig.~\ref{fig8}(b). This means that for much of the time (apart from $t=0,T/2$), the zero-energy Floquet mode populates even-numbered sites (from the left edge) and their symmetric reflection sites. This finding aligns with the recovery of chiral symmetry only at \(t=0\) or \(t=T/2\), as observed in Fig.~\ref{fig6}.
	
	To make the explanation more intuitive, we plot the modulus of the quasienergy as a function of $\gamma$, for defect pairs introduced at $m_0=2$ and its reflection-symmetric site, as shown in Fig.~\ref{fig8}(c) (high-frequency regime) and Fig.~\ref{fig8}(d) (low-frequency regime), where the other parameters are the same as in Fig.~\ref{fig8}(a) and Fig.~\ref{fig8}(b). 
	In the high-frequency regime, this zero-energy eigenstate exhibits no weight at the defect sites at any time, thus it does not participate in the $\mathcal{PT}$ symmetry breaking. Therefore, the zero-energy topological edge states are immune to the defects, preserving their profile and keeping their quasienergy at zero (marked in red), even beyond the $\mathcal{PT}$ symmetry breaking phase transition.
	In low-frequency regimes, the zero-energy Floquet mode of the defect-free system has no population at even-numbered sites on the left half of the lattice and their reflection-symmetric counterparts exclusively at \( t = 0 \) or \( T/2 \) (where chiral symmetry is recovered). At all other times, substantial population emerges at these sites for the zero-energy Floquet mode.
	Therefore, this zero-energy topological edge state participates in the $\mathcal{PT}$ symmetry breaking, resulting in a vanishing $\mathcal{PT}$ symmetry breaking threshold. As illustrated in Fig.~\ref{fig8}(d), the zero-energy topological edge state is affected by the defects,  which leads to a deviation of its quasienergy from zero.

	\section{SYMMETRY BREAKING Threshold in Odd-Sized Lattices: The Role of Topological Edge States}
\begin{figure}[htbp]
		\centering
		\includegraphics[trim=0 0 0 0,clip,width=9.2cm]{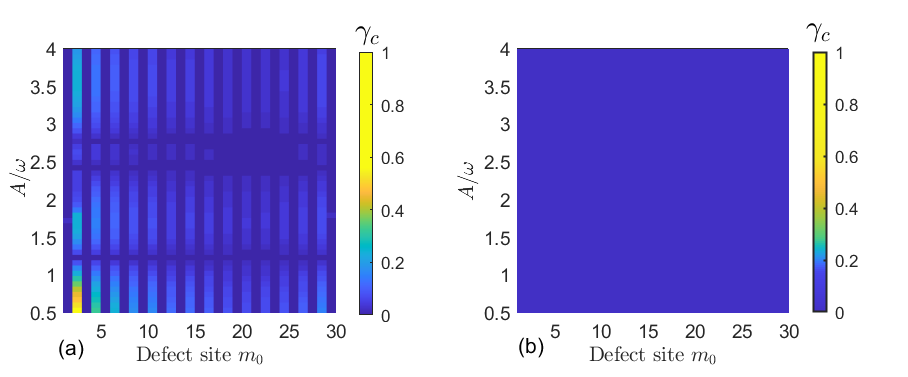}
		\caption{Symmetry breaking threshold versus $A/\omega$ and $m_0$ in the odd-sized lattice for (a) $\omega = 20$ (high-frequency) and (b) $\omega = 1$ (low-frequency). Parameters: $N=61$, $J=1$, $a=1$ and $b=2$.}
		\label{fig9}
	\end{figure}

\begin{figure}[htbp]
		\centering
		\includegraphics[trim=0 0 0 0,clip,width=9.2cm]{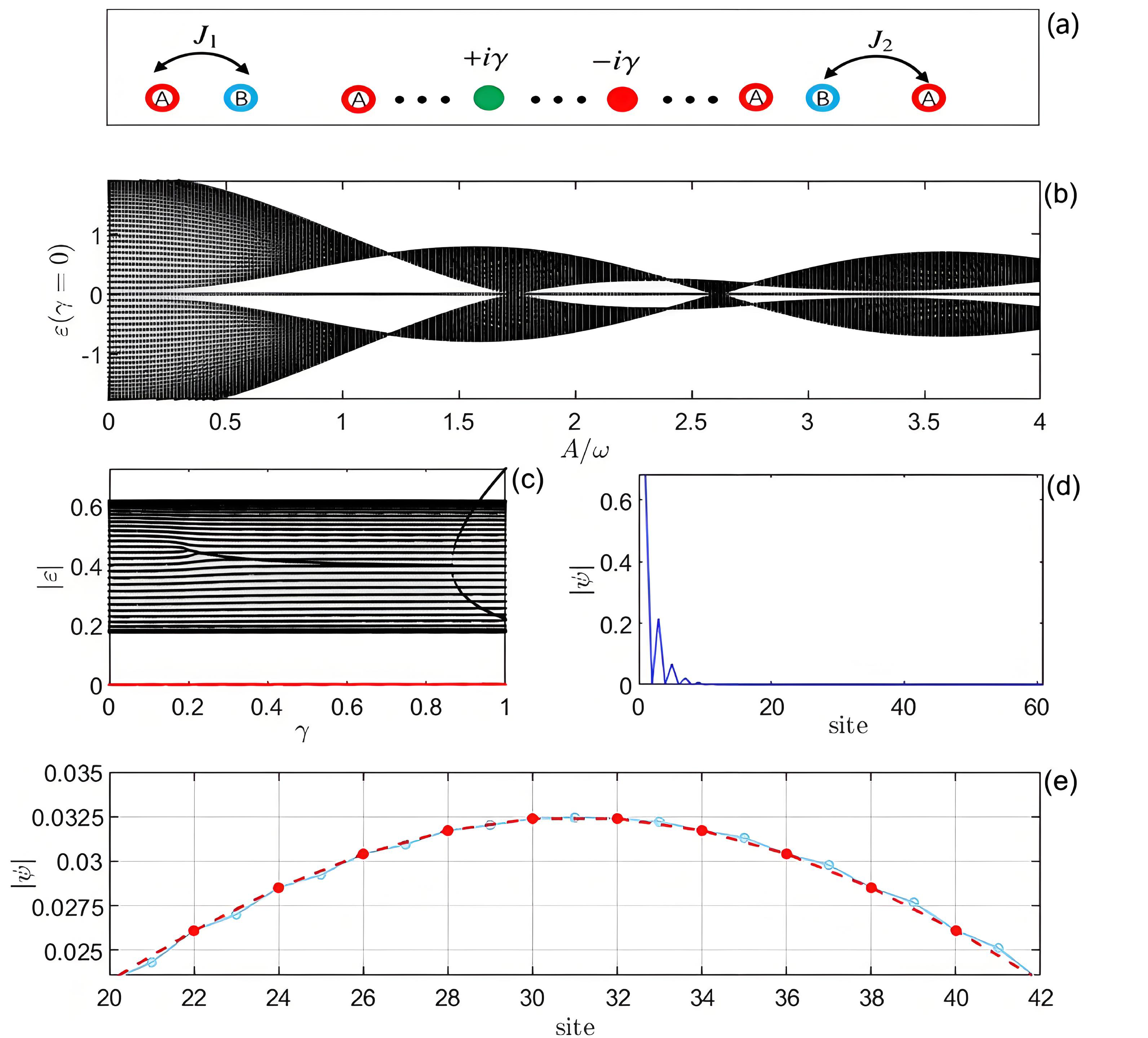}
		\caption{(a) Schematic of an odd-sized ($N$-site) chain with a bipartite lattice geometry. Balanced gain and loss are applied at a pair of reflection-symmetric sites, $m_0$ and $N+1-m_0$. In the high-frequency regime, the system is described by effective intra-cell ($J_1$) and inter-cell ($J_2$) hoppings [see Eq.~\eqref{con:7}]. (b) Numerically calculated quasi-energy spectrum for an open chain of $N=61$ sites in the absence of defect pairs at a high frequency $\omega = 20$. (c) Quasi-energy spectrum with $\mathcal{PT}$-symmetric defects introduced at $m_0=2$ and $N+1-m_0=60$. The red line denotes the topologically protected zero-energy bound state, whose energy is robust against the introduction of these defects. (d) Spatial profile of one topologically protected bound state in the symmetry broken phase, with parameters $m_0=2$ and $\gamma=0.4$. (e) Ground-state wavefunction of model~\eqref{con:1} in the high-frequency regime, revealing a hidden symmetry. The parameters are $A/\omega = 2$, $\gamma=0.4$ and $m_0=2$, with others identical to (b). The red dots (cyan circles) denote the population on even (odd) sites, demonstrating a clear reflection symmetry of the even-site amplitudes about the center of the chain. Parameters used are $N=61$, $\omega=20$, $J=1$, $a=1$ and $b=2$.}
		\label{fig10}
	\end{figure}

	In our work, we primarily focus on even-sized lattice systems. In this section, however, we consider an odd-sized lattice system, which consists of a periodic dimer sequence $ABAB...ABA$, as illustrated in Fig.~\ref{fig10}(a).
	In Fig.~\ref{fig9}, we show the $\mathcal{PT}$ symmetry breaking threshold strength $\gamma_c$ as a function of $m_0$ and $A/\omega$ for a lattice with $N = 61$ sites. Results for the high-frequency case ($\omega = 20$) are presented in Fig.~\ref{fig9}(a), while those for the low-frequency case ($\omega = 1$) are shown in Fig.~\ref{fig9}(b).
	As shown in Fig.~\ref{fig9}(a), the $\mathcal{PT}$ symmetry breaking threshold strength $\gamma_c$ exhibits an alternating pattern of zero and non-zero values across all $m_0$ for any value of $A/\omega$. This periodic behavior is distinct from what is observed in even-sized lattices, where such a repeating pattern typically exists only for specific range of $A/\omega$.
	We also observe that at specific driving amplitudes, $A/\omega = 1.2, 2.4, 2.8$, the $\mathcal{PT}$ symmetry breaking threshold strength $\gamma_c$ is zero, irrespective of the defect location $m_0$. This behavior arises from the occurrence of coherent destruction of tunneling (CDT), where intra-cell or inter-cell hopping is frozen.
	However, as Fig.~\ref{fig9}(b) indicates, in the low-frequency regime, the $\mathcal{PT}$ symmetry breaking threshold strength remains almost zero irrespective of where the defect is placed and irrespective of the magnitude of the driving parameter.
	
	In Fig.~\ref{fig10}(b), we show the calculated quasienergy spectrum for an open $N=61$ lattice in the high-frequency regime ($\omega = 20$), specifically for the defect-free case.
	As a result of the odd-sized lattice sites, the system exhibits zero-energy topological edge states throughout the entire range of the driving parameter.
	This accounts for the observed repeating pattern of the $\mathcal{PT}$ symmetry breaking threshold at all $A/\omega$ values, as shown in Fig.~\ref{fig9}(a).
	Next, we investigate the influence of zero quasi-energy topological edge states on the $\mathcal{PT}$ symmetry breaking threshold in the odd-sized lattice. Specifically, Figs. \ref{fig10}(c)-(e) examine the configuration with balanced gain and loss applied at reflection-symmetric sites $m_0 = 2$ and $N - m_0 + 1 = 60$ for the driving parameter $A/\omega = 2$.
	In Fig.~\ref{fig10}(c), we observe that the zero-energy topological edge states are unaffected by the defect strength, keeping their quasienergy constantly at zero (highlighted in red). 
	When $A/\omega = 2$, the effective hopping strength is adjusted to $J_1 < J_2$ [see Eq.~\eqref{con:7}]. As a result, the system has only one zero-energy edge state localized on the left side of the lattice [see Fig.~\ref{fig10}(d)], not populating any even-numbered sites, and thus having no effect on the $\mathcal{PT}$ symmetry breaking threshold.
	If the parameter \( A/\omega \) is adjusted such that \( J_1 > J_2 \), the system hosts only one zero-energy edge state, localized on the right side of the lattice, also without affecting the $\mathcal{PT}$ symmetry breaking threshold.
	
	As observed from the schematic in Fig.~\ref{fig10}(a), the system is generally not invariant under $\mathcal{PT}$ operation. This raises the question: why is there still a positive $\mathcal{PT}$ symmetry breaking threshold?
	In Fig.~\ref{fig10}(e), we examined the profile of the ground state before symmetry breaking. 
	As expected, the ground-state profile shows reflection asymmetry about the central site \( n = 31 \). Nevertheless, the wave function possesses a hidden symmetry~\cite{harter2016mathcal}: the amplitudes on even sites (red dots) are spatially inversion symmetric, while those on odd sites (cyan circle) are not.
	This explains the non-zero $\mathcal{PT}$ symmetry breaking threshold observed in Fig. ~\ref{fig9}(a) when $m_0$ is even. The hidden symmetry, reminiscent of that in spatially asymmetric, time-dependent Aubry-André and Harper models~\cite{harter2016mathcal}, emerges in our system under high-frequency driving. This symmetry breaks down at low frequencies, resulting in the vanishing of the symmetry-breaking threshold [see Fig.~\ref{fig9}(b)].

	\section{Control of the $\mathcal{PT}$-SYMMETRY BREAKING Threshold by Two Driving Fields}
	\begin{figure*}[htbp]
		\centering
		\includegraphics[trim=0 0 0 0,clip,width=14cm]{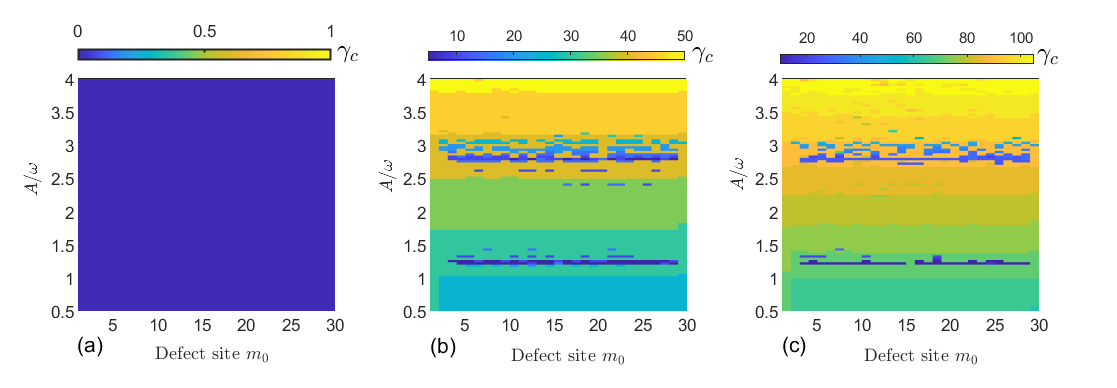}
		\caption{$\mathcal{PT}$ symmetry breaking threshold in an even-sized lattice under the combined periodic driving defined in Eq.~\eqref{con:20}, shown for various phase shifts \(\theta\)  and driving frequencies \(\omega\). (a) \(\theta = \pi/2\), \(\omega = 20\). (b) \(\theta = 0\), \(\omega = 10\). (c) \(\theta = 0\), \(\omega = 20\).  The other parameters are \(N=60\), \(J=1\), \(a=1\), and \(b=2\).}
		\label{fig11}
	\end{figure*}
	Furthermore, we add a co-frequency driving term \(\cos(\omega t-\theta)\) to the \(\mathcal{PT}\)-symmetric defects, and return the model to the even-sized lattice, for example, \(N=60\) in our numerical simulation. To distinguish this model from Eq.~\eqref{con:1}, we denote the Hamiltonian subjected to the combined periodic driving as $H_{d}(t)$, which is given by
	\begin{equation}\label{con:20}
		\begin{split}
			H_{d}(t) &= \sum_{n}^{N} J(\left| n \right\rangle\langle n+1 |+\left| n+1 \right\rangle\langle n |)+A\sin(\omega t)\sum_{n}^{N} x_{n}\left| n \right\rangle\langle n |\\
			&+ i\gamma \cos(\omega t-\theta)\left| m_0 \right\rangle\langle  m_0 |\\ 
			&- i\gamma \cos(\omega t-\theta)\left| N-m_0+1 \right\rangle\langle  N-m_0+1 |.
		\end{split}
	\end{equation}
	For \(\theta = 0\), there exists a time instant \(t_0 = \pi/(2\omega)\) satisfying the time-reversal symmetry condition: 
	\begin{equation}\label{con:21} 
		{H_{d}}^{*}(t_{0} + t) = H_{d}(t_{0} - t). 
	\end{equation} 
	However, for \(\theta=\pi/2\), time-reversal symmetry is violated, and no such \(t_0\) exists. Symmetry analysis predicts that when the Hamiltonian satisfies time-reversal symmetry, there exists the positive \(\mathcal{PT}\) symmetry breaking threshold, and when time-reversal symmetry is broken, there is no \(\mathcal{PT}\) symmetry breaking threshold.
	This symmetry analysis prediction is confirmed by our numerical simulation, as shown in Fig.~\ref{fig11}, where we consider the high-frequency case.
	As shown in Figs.~\ref{fig11}(b) and (c), when \(\theta = 0\) is chosen such that the Hamiltonian satisfies time-reversal symmetry, there exists a non-zero \(\mathcal{PT}\) symmetry breaking threshold. In contrast, when \(\theta = \pi/2\) is chosen, time-reversal symmetry is broken, and the threshold becomes zero [Fig.~\ref{fig11}(a)].
	Additionally, we note that double driving can make the \(\mathcal{PT}\) symmetry breaking threshold very large compared to the single driving case, and this threshold increases as the driving frequency increases.
	
	These observations can be systematically explained through high-frequency asymptotic analysis. In the following, we consider the case of $\theta = 0$ as an example, which ensures the Hamiltonian \eqref{con:20} is time-reversal symmetric. Following the same method as in Sec.~$\textcolor{blue}{\mathrm{III}}$, we first apply a non-unitary similarity transformation 
	\begin{equation}\label{con:22} 
		\begin{split} \hat{S}(t) &= \text{exp} \Biggl\{-i\int \Biggl[A\sin(\omega t)\sum_{n}^{ N} x_{n}\left| n \right\rangle\langle n | + i\gamma \cos(\omega t) \vert m_0 \rangle \langle m_0 \vert \\ & - i\gamma \cos(\omega t) \vert N - m_0 + 1 \rangle \langle N - m_0 + 1 \vert \Biggr]dt\Biggr\}.
		\end{split} 
	\end{equation}
	The Hamiltonian after rotation is given by, ${H}_d'(t)=\hat{S}^{-1 } (t) H_{d}(t) \hat{S} (t)-i\hat{S}^{-1 } (t)\frac{\mathrm{d} \hat{S}(t) }{\mathrm{d} t} $, leading to
	\begin{equation}\label{con:23}
		\begin{split}
			{H}_d'(t) & = \sum_{n}^{N-1} \left( K_{n} |n\rangle\langle n+1| + K_{n}^{-1} |n+1\rangle\langle n| \right),
		\end{split}
	\end{equation}
	where
	\begin{align*}
		K_{n} & =
		\begin{cases}
			t_n & n \neq m_0-1, N-m_0, N-m_0+1, m_0 \\
			t_n e^{\gamma \sin(\omega t)/\omega} & n = m_0-1, N-m_0+1 \\
			t_n e^{-\gamma \sin(\omega t)/\omega} & n = m_0, N-m_0
		\end{cases} \\
		\text{with} \quad & \\
		t_n & = J e^{i\frac{A}{\omega}\cos(\omega t)(x_{n+1}-x_n)}.
	\end{align*}
	It should be noted that the transformed Hamiltonian in Eq.~\eqref{con:23} is still non-Hermitian.
	Given that $\omega$ is much larger than all the
	other energy scales of the problem, the quantum evolution of this system can be governed by the time-independent effective Hamiltonian, which is derived from a $1/\omega$ expansion and is expressed as
	$\hat{H}_{\text{eff}}=\hat{H}_{\text{eff}}^{(0)}+\frac{1}{\omega }\hat{H}_{\text{eff}}^{(1)}+\frac{1}{\omega^2 }\hat{H}_{\text{eff}}^{(2)}+...$.
	To the lowest order of the $1/\omega$ expansion, we keep the zero-frequency component and obtain
	\begin{equation}\label{con:24} 
		\begin{split} {H}_\text{{eff}}^{(0)}&=\frac{1}{T} \int_{0}^{T} dt{{H}'_{d} }(t)\\ &=\sum_{n}^{N-1}{K}^{\prime}_{n} \left|n\right\rangle\left\langle n+1 \right|+{K}^{\prime}_{n}\left|n+1\right\rangle\left\langle n \right|, 
		\end{split} 
	\end{equation} 
	where
	\begin{widetext}
		\begin{align*} 
			{K}^{\prime}_{n} = \begin{cases} J \mathcal{J}_{0}[\frac{A}{\omega}(x_{n+1}-x_{n})]& n \neq m_0-1, N-m_0,N-m_0+1,m_0, \\ J\sum_{l=-\infty}^{\infty}(-1)^{l}\mathcal{J}_{l}[\frac{A}{\omega}(x_{n+1}-x_{n})] I_{l}(\frac{\gamma}{\omega}) & n = m_0-1, N-m_0+1, \\ J\sum_{l=-\infty}^{\infty}\mathcal{J}_{l}[\frac{A}{\omega}(x_{n+1}-x_{n})] I_{l}(\frac{\gamma}{\omega}) & n = m_0, N-m_0.  \end{cases} 
		\end{align*} 
	\end{widetext}
	Here, \(I_l(x)\) denotes the modified Bessel function of the first kind, which is related to the Bessel function of the first kind, \(\mathcal{J}_l(x)\), by the identity \(\mathcal{J}_l(-ix) = (-i)^l I_l(x)\).
	
	Notably, \(I_l(x)\) is real-valued, which ensures that the zeroth-order approximation of the effective Hamiltonian is Hermitian and therefore possesses a real energy spectrum. This explains why the system's \(\mathcal{PT}\) symmetry breaking threshold is significantly increased when \(\theta=0\) in the high-frequency limit.
	However, since the full Hamiltonian is non-Hermitian, higher-order corrections to this approximation will introduce complex energies. Furthermore, as indicated by the expansion of the effective Hamiltonian, the influence of these higher-order terms diminishes at higher driving frequencies. This causes the \(\mathcal{PT}\) symmetry breaking threshold to increase with frequency, as shown in Figs.~\ref{fig11}(b) and (c). This behavior indicates that for the case of double driving, the system effectively behaves more like a Hermitian one at higher frequencies.

	\section{CONCLUSIONS}
	We have systematically investigated the control mechanism of the \(\mathcal{PT}\)-symmetry breaking threshold in Floquet-driven bipartite lattices with spatially symmetric gain-loss defects, and have revealed the frequency-dependent influence of topological edge states on the threshold.
	In the high-frequency regime, once the driving field has induced a topological phase, the \(\mathcal{PT}\)-symmetry breaking threshold is critically determined by the position of the defect.
	When the defect is located where the edge state has nonvanishing components, the system's \(\mathcal{PT}\) symmetry is immediately broken, resulting in a symmetry-breaking threshold of zero.
	Conversely, when the defect is placed where the edge state has no population—specifically, on a site where the edge state's wavefunction component is zero (e.g., an even-numbered site from the left edge or its reflection-symmetric counterpart)—the threshold is then governed by defect-induced bound states.
	In the low-frequency driving regime, the active participation of Floquet-induced topological edge states, characterized by their dynamic spatial evolution over one driving period, universally reduces the symmetry-breaking threshold to zero, irrespective of where the defect is placed.
	
	Extending our study to lattices of odd size ($N$ odd), we uncover a distinct high-frequency behavior in contrast to even-sized systems: the symmetry-breaking threshold, $\gamma_c$, alternates between zero and non-zero as a function of the defect location for all driving amplitudes $A/\omega$. This striking pattern is attributed to the presence of a zero-energy topological edge state that emerges for all driving parameters. Conversely, the low-frequency regime consistently induces a universal suppression of symmetry-breaking threshold. We explain this phenomenon by tracing its origin to a hidden symmetry associated with the even-site occupation in the system's Floquet eigenstates.
	We further demonstrate that applying a co-frequency drive to the \(\mathcal{PT}\)-symmetric defect introduces a powerful new control knob. Under preserved time-reversal symmetry, this protocol renders the the high-frequency zeroth-order effective Hamiltonian  Hermitian, thereby dramatically increasing the symmetry-breaking threshold.
	The mechanisms we elucidate—namely, the stark contrast in the influence of topological edge states between high- and low-frequency regimes, the universal suppression of the threshold at low frequencies, and the tunability via combined modulation of the lattice and the defect pairs—provide a new framework for manipulating non-Hermitian phenomena.

	\section*{ACKNOWLEDGMENTS}
	The work was supported by the National Natural Science Foundation of China (Grants No. 12375022, No. 11975110 and No.12275033), the Natural Science Foundation of Zhejiang Province (Grant No. LY21A050002), Zhejiang Sci-Tech University Scientific Research Start-up Fund(Grant No.
	20062318-Y), Jiangxi Provincial Natural Science Foundation(No. 20232BAB201008), and Education Department of Jiangxi Province (GJJ2201629).

	\section*{APPENDIX}
	\setcounter{equation}{0}
	\renewcommand\theequation{A\arabic{equation}} 
	The model \eqref{con:1} can also be realized in a photonic waveguide array, such as the one depicted in Figs.~\ref{fig1}(c)-(d). In this system, waveguides arranged in alternating upper and lower rows are periodically bent along the longitudinal propagation direction ($z$). It is well-established that the propagation of light, specifically the slowly-varying electric field amplitude $\psi(x,y,z)$, is governed by the paraxial Schr\"odinger-like wave equation:
	\begin{equation}\label{con:101}
		i\frac{\partial\psi}{\partial z}=-\frac{1}{2}\left(\frac{\partial^2}{\partial x^2}+\frac{\partial^2}{\partial y^2}\right)\psi+{V}\left(x,y,z\right)\psi.
	\end{equation}
	Here, $x$, $y$, and $z$ are normalized coordinates, and $V(x,y,z)$ represents the refractive index landscape of the waveguide array. This potential is modeled as a sum of Gaussian functions:
	\begin{equation}\label{con:102}
		\begin{aligned}
			V(x,y,z)&=\sum_{n=1}^{N}V[x-\tilde{x}_{n}(z),y-\tilde{y}_{n}] \\& =\sum_{n=1}^{N}p  \text{exp}{\left\{\left[-\frac{x-\tilde{x}_{n}(z) }{d}\right]^{6}+\left[-\frac{y-\tilde{y}_{n}}{d}\right]^{6}\right\} }\\&+ip^{\prime}\text{exp}{\left\{\left[-\frac{x-\tilde{x}_{m_0}(z) }{d}\right]^{6}+\left[-\frac{y-\tilde{y}_{m_0} }{d}\right]^{6} \right\}}\\&-ip^{\prime} \text{exp}{\left\{\left[-\frac{x-\tilde{x}_{N-{m_0}+1}(z) }{d }\right]^{6}+\left[-\frac{y-\tilde{y}_{N-{m_0}+1}}{d}\right]^{6} \right\}}. \\
		\end{aligned}
	\end{equation}
	The horizontal center position of the $n$-th waveguide varies sinusoidally with $z$ according to $\tilde{x}_{n}(z)=\tilde{x}_{n}(0)+A\sin(\omega z)$, where $\tilde{x}_{n}(0)$ is its initial position at $z=0$. 
	This periodic bending, illustrated in Fig.~\ref{fig1}(d), is applied only horizontally; the vertical positions $\tilde{y}_{n}$ are constant, taking on one of two fixed values corresponding to the upper and lower rows.
	To implement the $\mathcal{PT}$-symmetric potential, specific waveguides at sites $m_0$ and $N-m_0+1$ are engineered with gain (blue) and loss (yellow), as highlighted in Figs.~\ref{fig1}(c)-(d).
	The parameters $p$ and $p'$ describe the refractive index contrast of their respective waveguides, and $d$ denotes the waveguide width.
	
	To simplify the analysis, we transform to a co-moving coordinate system $(x^{\prime},y^{\prime},z^{\prime})$ defined by $x^{\prime}=x-\tilde{x}_{n}(0)-A\sin(\omega z)$, $y^{\prime}=y$, and $z^{\prime}=z$. Applying the similarity transformation \(\phi(x^{\prime}, y^{\prime},z^{\prime})=\psi(x^{\prime}, y^{\prime},z^{\prime})\text{exp}\left(\frac{i\dot{x}^{\prime}x^{\prime}}{2}\right)\text{exp}\left[-i\int \left(\frac{\dot{x}^{\prime}}{2}\right)^2dz^{\prime}\right]\), the wave equation~\eqref{con:101} transforms into
	\begin{equation}\label{con:103}
		\begin{aligned}
			i\frac{\partial}{\partial z^{\prime}}\phi(x^{\prime},y^{\prime},z^{\prime})=&-\frac{1}{2}\left(\frac{\partial^2}{\partial x^{\prime2}}+\frac{\partial^2}{\partial y^{\prime2}}\right)\phi(x^{\prime},y^{\prime},z^{\prime})\\&	+{V}\left(x^{\prime},y^{\prime}\right)\phi(x^{\prime},y^{\prime},z^{\prime})-\frac{1}{2}\ddot{x}^{\prime}x^{\prime}\phi(x^{\prime},y^{\prime},z^{\prime}),
		\end{aligned}
	\end{equation}
	where the potential $V(x^{\prime},y^{\prime})$ is now time-independent and given by
	\begin{align*}
		{V}\left(x^{\prime},y^{\prime}\right)&=\sum_{n=1}^{N}p \ \text{exp}{\left\{\left[-\frac{x^\prime-\tilde{x}_{n}(0) }{d}\right]^{6}+\left[-\frac{y^\prime-\tilde{y}_{n} }{d}\right]^{6} \right\}}\\&+ip^{\prime}\text{exp}{\left\{\left[-\frac{x^\prime-\tilde{x}_{m_0}(0) }{d}\right]^{6}+\left[-\frac{y^\prime-\tilde{y}_{m_0}}{d}\right]^{6} \right\}}\\&-ip^{\prime} \text{exp}{\left\{\left[-\frac{x^\prime-\tilde{x}_{N-{m_0}+1}(0) }{d}\right]^{6}+\left[-\frac{y^\prime-\tilde{y}_{N-{m_0}+1}}{d}\right]^{6} \right\}}.
	\end{align*}
	In the above, $\dot{x}^{\prime}$ and $\ddot{x}^{\prime}$ denote the first and second derivatives of $x^{\prime}$ with respect to $z$, respectively.

	We now expand the field into a superposition of the individual waveguide modes:
	\begin{equation}\label{con:104}
		\phi(x^{\prime},y^{\prime},z^{\prime})=\sum_n^{N}\varphi_n(z)w_n(x,y),
	\end{equation}
	which yields the following coupled-mode equations:
	\begin{equation}\label{con:105}
		\begin{aligned}
			-i\frac{d\varphi_n}{dz} & =\tau_n\varphi_{n+1}+\tau_{n-1}\varphi_{n-1}+D_n\varphi_n,
		\end{aligned}
	\end{equation}
	where the coupling coefficient $\tau_{n}$ and the on-site potential $D_{n}$ are defined as
	\begin{equation*}
		\begin{aligned}
			\tau_{n}&=\iint w_{n}^{*}(x^\prime,y^\prime)\Biggl[-\frac{1}{2}\left(\frac{\partial^2}{\partial x^{\prime2}}+\frac{\partial^2}{\partial y^{\prime2}}\right)\\&+V(x^{\prime},y^{\prime})\Biggr]w_{n+1}(x^\prime,y^\prime)dx^\prime dy^\prime,\\
			D_{n}&=\iint w_{n}^{*}(x^{\prime},y^\prime)\Biggl[V(x^{\prime},y^{\prime})-\frac{1}{2}\ddot{x}^{\prime}x^{\prime}\Biggr]w_{n}(x^\prime,y^\prime)dx^\prime dy^\prime\\&=\begin{cases}
				A\sin(\omega t)\tilde{x}_{n}(0) &  n \neq m_0, N-m_0+1, \\
				A\sin(\omega t)\tilde{x}_{n}(0)+i\gamma &  n = m_0, \\
				A\sin(\omega t)\tilde{x}_{n}(0)-i\gamma &  n =  N-m_0+1.
			\end{cases}
		\end{aligned}
	\end{equation*}
	This derivation shows that the periodic bending is equivalent to an additional oscillating on-site potential in the Hamiltonian. 
	The Hamiltonian for the system described by the coupled-mode equations~\eqref{con:105} is the same as that in Eq.~\eqref{con:1}.

\end{document}